# Field Emission properties of nanocomposites of conducting polymers

Pejman Talemi
School of Chemical Engineering, University of Adelaide, SA 5005, Australia
Email: pejman.talemi@adelaide.edu.au

## Introduction

Field emission is the process of emission of electrons from a clod cathode upon application of an external electric field. This process has found many applications in future display devices, X-Ray generators, and electron microscopes[1-3]. Carbon-based nanomaterials such as carbon nanotubes (CNTs)[4-9] or graphene[9-12] are shown to be good candidates for development of cold cathodes for field emission process. However, most of these materials are in powder form and development of methods for fabricating stable macroscale electrodes seems inevitable for production of field emission devices. Preparation of composites of these particles with polymers[13-14], metals[15] or conducting polymers[16-19] appears to be a feasible strategy for development of macroscale cold cathodes. Amongst these materials, conducting polymers seems to be able to combine the advantages of metals and polymers by providing a good level of conductivity, good adhesion and interaction with carbon nanomaterials and the ease of processing. However, as we have shown in a previous work[15] (in the Supplementary Materials section) due to the lower thermal stability and electrical conductivity, these polymers cannot provide the required stability for functioning of a field emission cathode over long period of time or at high current densities.

In this report, we will provide additional data and discussion on performance of these electrodes in field emission devices. We will also evaluate some strategies for improving the field emission performance of these composites. In this work composites of polypyrrole and carbon nanotubes were prepared using two different methods. In the first method, the chemical oxidative polymerization of pyrrole was used for preparation of a composite powder of carbon nanotubes and polypyrrole which was later pressed into a solid disk (Fig 1a) and used for field emission studies. The second method for preparation of polypyrrole/carbon nanotube nanocomposite was based on electro-polymerization. In this process polymerization of pyrrole in presence of an anionic surfactant and carbon nanotubes were tried. It is shown that such polymerization process can lead to the formation of columnar composite structures[20]. At the end, in order to overcome problems with conducting polymers we attempted to electropolymerize the polymer over an insulating membrane[21]. This membrane can be used for controlling field emission by limiting the emitting area which can also reduce the ohmic heat generation in the system and stabilize the field emission. It is expected that the membrane can also help the heat dissipation from the electrode.

## Experimental

*Preparation of polypyrrole/carbon nanotube nanocomposite*

polypyrrole/ carbon nanotube nanocomposite particles were prepared by oxidative polymerization of pyrrole under sonication, using a method previously reported for synthesis of iron oxide/PPy nanocomposites[22]. 0.2 g of pyrrole (Sigma-Aldrich, reagent grade) and 0.02 g of carbon nanotubes (Nanocyl 3150, Nanocyl S.A, Belgium) were added to 100 ml of deionized water and sonicated for 10 minutes in an ice bath using a SonicsVibraCell ultrasonic wand. In the next step the ammonium persulfate (Sigma-Aldrich) solution in deionized water were added to the mixture drop-wise, under vigorous mechanical stirring. The resultant composite was separated by centrifugation, and was washed with deionized water several times.

*Preparation of polypyrrole/graphene nanocomposites*

Graphene solutions were prepared by oxidation-exfoliation-reduction method from natural graphite (SP-1, Bay Carbon), using the recipe reported by Li et al [23]. 100 ml of the solution containing 20mg of graphene were mixed with 0.2g of pyrrole, the ammonium persulfate solution added to the mixture drop-wise under vigorous stirring using a mechanical stirrer. The prepared composite were separated by centrifuging and was washed with deionized water for several times.

*Preparation of polypyrrole/carbon nanotube nanocomposite by electropolymerization*

A solution of 0.1 mol/L sodium dodecyl sulphate, 25mmole/L pyrrole and 0.17g/L carbon nanotubes was prepared in deionized water. The electropolymerisation process was performed in a standard three-electrode cell, using a Faraday M1 Obligato potatiostat, with a cycling voltage mode with scan rate of 50mV/s in a potential range of -0.2 and +0.8 V vs. saturated calomel electrode. The working electrode in this experiment was a gold-coated stainless steel electrode. The prepared film was then washed with deionized water.

*Electropolymerization of CNT/PPY composites over a membrane*

Electropolymerization of CNT/PPY over a membrane was performed following the method reported before[21]. The anodic aluminum oxide (AAO) membrane (Anodisc 25, pore size 100nm, Whatman, USA) was adhered to the surface of a copper electrode and sealed, so that the only contact between the copper electrode and solution was through the membrane pores. The electro-polymerization process was performed in a standard three-electrode cell, a constant potential of +0.8 V versus SCE was applied using a Faraday M1 Obligato potatiostat, for 3 minutes. The film thus prepared was washed with deionized water and characterized.

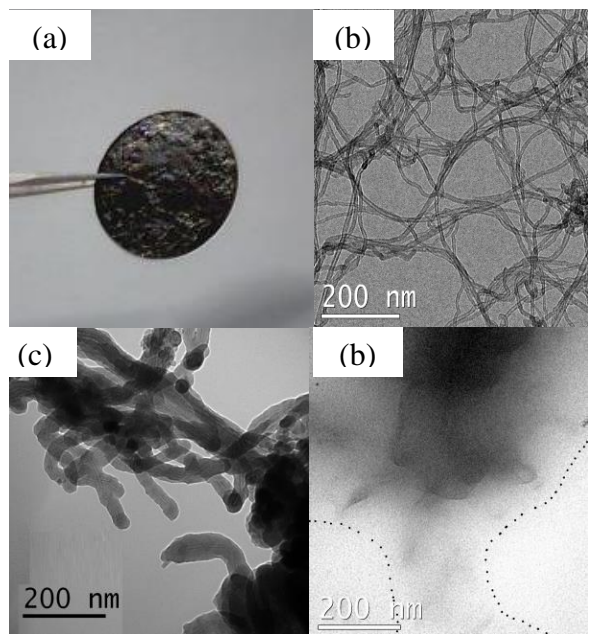

**Fig. 1** (a) The composite sample pressed into a disk for field emission measurements and (b) TEM images showing the raw carbon nanotubes, (c) formation of a layer of polypyrrole around nanotubes and (d) the graphene sheets embedded inside polypyrrole particles. The dotted line is used to show the boundary of the composite particle.

**Results and Discussion**

*Preparation and characterization of composites*

The TEM images of the powder sample confirm that this process has resulted in formation of a layer of polypyrrole around each carbon nanotube (Fig. 1b). In case of Graphene/polypyrrole composites, TEM images of samples confirms the presence of graphene in form of folded sheets embedded inside the polypyrrole particles (Fig. 1d), approving the formation of the desired composite particles.

The thermogravimetric analysis of these nanocomposites in air with a scan rate of 10°C/min demonstrates a two-step degradation pattern (Fig.2). The first step commencing at around 200°C, and is related to oxidation of polypyrrole. The second weight loss occurs at around 600 °C and is ascribed to combustion of graphene nanosheets. The evaluation of composition of these two components using TGA graphs shows that although the concentration of graphene in the initial mixing process was 10 wt. %, the final composite consists of ~17 wt. % graphene which means that some of the added pyrrole has been washed away without contributing in formation of the final composite.

Poly(3,4-ethylenedioxythiophene) (PEDOT) is another member of the family of conducting polymers, and is known to exhibit better conductivity, flexibility[24], optical properties[25] and thermal stability in comparison to polypyrrole[26-29]. In order to test if the higher thermal stability of PEDOT, compared to PPy, can provide a better field emission performance, samples of PEDOT/CNT composites were prepared using the same methodology. Fig. 2 shows an SEM image of PEDOT/CNT composite. In the same way as other samples the composite powder was pressed into disks and used for electron field emission measurements.

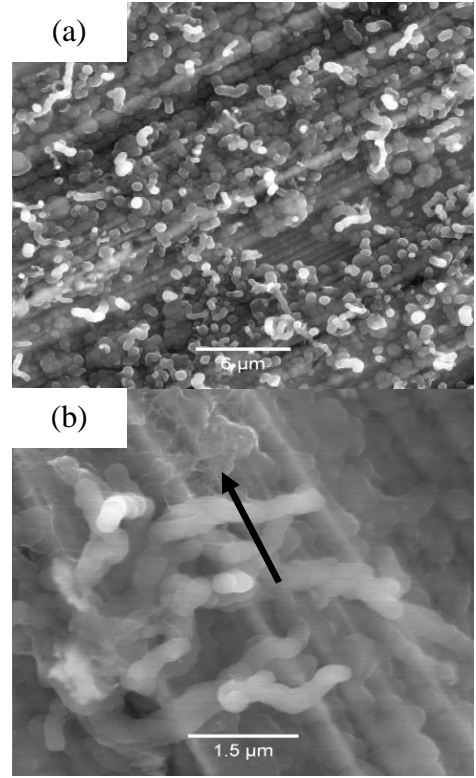

**Fig. 3** SEM images showing the surface of (a) the sample prepared by 10 CV cycles, and (b) the sample after 2 CV cycles, the arrow shows an area of carbon nanotubes with no polypyrrole coating.

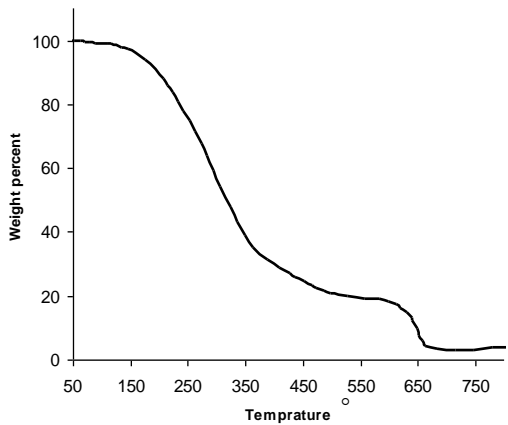

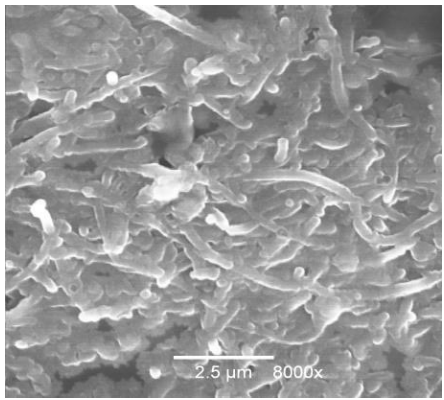

**Fig. 2** Thermogravimetric analysis of PPy/CNT (top) and SEM image (bottom) of a PEDOT/CNT composite.

Electropolymerization of conductive polymers has been shown to be a good method for the deposition of these polymers, and the resulting composites have better properties such as higher conductivity or improved chemical stability[30-31]. Electropolymerization is also shown to be able to generate a wide range of nanostructures on the surface[32-33]. Interestingly, the electropolymerization process described by Zhang et el.[20] results in vertically aligned columns of the composite material over the surface that is an ideal structure for field emission. Following their recipe the electropolymerization process was performed by applying voltage cycles with a scan rate of 50mV/s in the range of -0.2 to 0.5 volts versus saturated calomel electrode. Two samples were prepared using this technique, one after 2 voltage cycles and one after 10 cycles. As can be observed in Fig 3a, this process can efficiently produce composite nanostructures. At longer process times (10 cycles), the surface is covered with such columnar structures and all of the carbon nanotubes are covered with polypyrrole. While at the shorter polymerization times (2 cycles), some uncoated carbon nanotubes can be observed on the surface (Fig. 3b), which indicates that the process of migration of carbon nanotube-containing micelles to the surface of electrode occurs prior to polymerization process. Shorter processing times result in less crowded areas of columns, which indicates that by increasing the processing time, the diameter of the resulting columns remains unchanged within the range of 100-500 nm, while longer processing times only results in formation of new columns of similar diameter, and ultimately a more crowded morphology.

*Field emission Performance of nanocomposites*

The field emission properties of all samples have been studied using a parallel plate setup at a pressure of lower than $10^{-4}$ Pa. A gold coated stainless steel was used as the anode and a polyimide film was used as the spacer between electrodes to maintain the distance at 500μm. The first nanocomposite samples examined were those of carbon nanotubes and polypyrrole (CNTPPy) and the composite of graphene and polypyrrole (GRPPy). Figure 4 shows the plot of the variations in the measured current density (J) as a function of applied electric field (E) for these two samples. It can be observed that the CNTPPy sample is exhibiting a stronger emission compared to the graphene based sample. This can be due to the whisker-like[34] shape of nanotubes and is in agreement with our previous observations[11]. The main problem observed during the field emission measurement of these samples is the low stability of these samples under the electric field. As can be observed in Figure 4 at the electric fields greater than 2.5 V/μm, the emission becomes very unstable. This instability can be related to the thermal degradation of polypyrrole due to resistive heat generation in the sample. The degradation of the polymer results in formation of some gases in the space between electrodes which interferes with electron emission process and increases the pressure in the test chamber which was observed as a sudden simultaneous change in the vacuum of the system. In a typical test under continuous operation of a turbo molecular vacuum pump the pressure of the chamber was stable at ~$5\times10^{-5}$ Pa. Upon application of an electric fields of higher than 2.5 V/μm, the pressure suddenly enhanced by 4 times and reached to ~$2\times10^{-4}$ and by stopping the voltage the vacuum came back to its initial level in a short time. The effect of this degradation can also be seen in the SEM images of the surface of these samples after field emission test. Figure 5 shows the SEM images of the surface of the GRPPy sample before and after field emission measurement. Before commencing the field emission measurement process and due to the pressing of the sample into a disk, the surface of the sample is flat with some micron-scale holes and cracks. However, the field emission tests appear to result in removal of polypyrrole from the surface, and some of the graphene sheets which previously were embedded inside the polymer can now be observed. Although this phenomenon can potentially enhance the field emission by providing protruding emission tips, continuation of the field emission process and further removal of polypyrrole which holds graphene sheets together will cause the graphene sheets to become loose and free to span the two electrodes, leading to a short circuit in the system and damage to the field emission device.

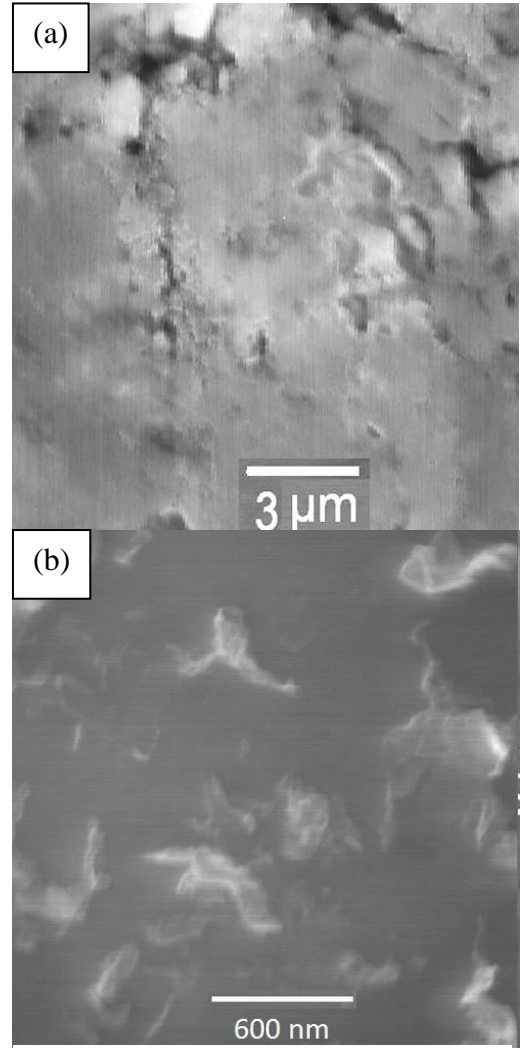

**Fig. 5** The SEM images of GRPPy sample before (a) and after (b) field emission test

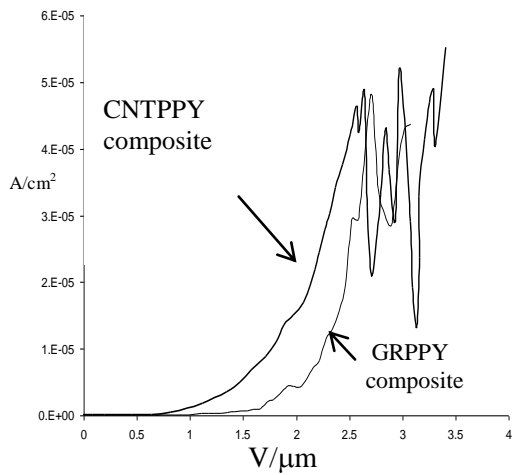

**Fig. 4** Current density vs. applied field graphs of CNTPPy and GRPPy.

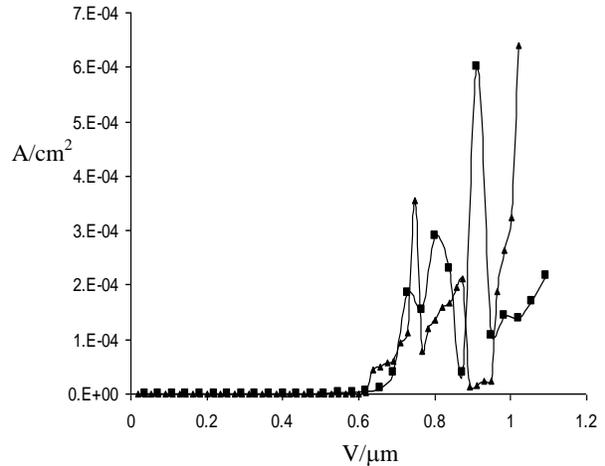

**Fig. 6** Current density vs. applied field graphs two of CNT/PEDOT samples with the same composition

As an approach for improving the longevity of conductive polymer composites samples upon application of an electrical field a polymer matrix with better conductivity and thermal stability has been employed for preparation of composites. PEDOT was chosen to be used in this regard. Figure 6 shows the field emission graph of a sample incorporating 10% of nanotubes in PEDOT via chemical polymerization. This graph shows the measurements performed on two different samples with the same composition. It can be seen that in comparison to CNTPPy sample, this sample demonstrates a lower turn-on voltage, and higher emission current but the emission is also extremely unstable and it seems that the nominally relatively higher stability of PEDOT, in comparison to polypyrrole [26-29], is not yet enough for providing a stable performance.

The results of field emission measurement of the nanocomposite samples prepared via electro-polymerization process using an anionic surfactant as counter ion, are shown in Figure 7. As discussed before, increasing the number of voltage cycles (polymerization time) results in larger number of fibers and a better coverage of the surface of electrode by the polypyrrole/carbon nanotube columnar nanocomposite structures. Figure 6 shows that the sample prepared by 10 cycles exhibits a better field emission and lower turn on voltage. These samples exhibit a higher emission in comparison to CNTPPy and GRPPy samples prepared by chemical polymerization, which indicates that the formation of aligned columnar composite nanostructures (Fig.3) in these samples can efficiently improve the field emission behavior. The shape of the field emission graphs of these samples also shows lots of fluctuations, mainly at electric field higher than 2.5V/μm which is again due to degradation of polypyrrole in the process. The range of the fluctuations in this sample is much narrower in comparison to the CNTPPy samples discussed before, which could be due to lower polypyrrole content in these samples.

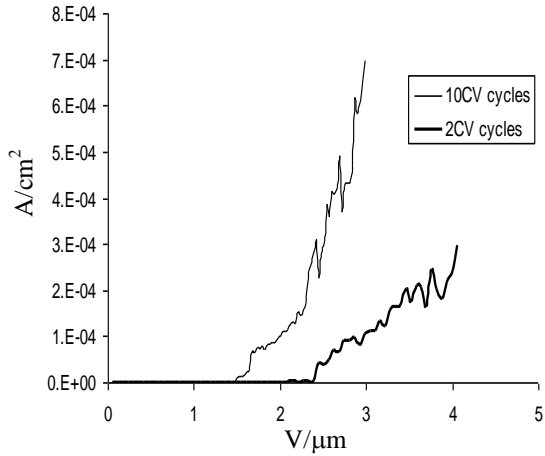

**Fig. 7** Current density vs. applied field plot of CNTPPy samples prepared by the electropolymerization process.

*Field emission performance of composites deposited over a membrane*

In order to control current density and to provide better heat transfer it was proposed to deposit the composite over a membrane. A stable porous ceramic membrane (anodized aluminum oxide (AAO)) can act as a spacer and provides a path through which the electrons can travel towards the anode in the field emission cell.

This membrane can be used to control field emission by limiting the emitting area which can also reduce the ohmic heat generation in the system and stabilize the field emission. However, the electropolymerization technique is limited to conductive substrates. Thus, it cannot be used for deposition of polymers over an insulating membrane. In order to overcome this problem we have used a recently reported technique[21] for direct electropolymerization of PPy/CNT nanocomposite on the surface of an electrically-nonconductive membrane. It is hypothesized that direct deposition of these composite materials facilitates the production process and provide a stable emission. The experimental details and the mechanism of this electropolymerization process is described in a previous

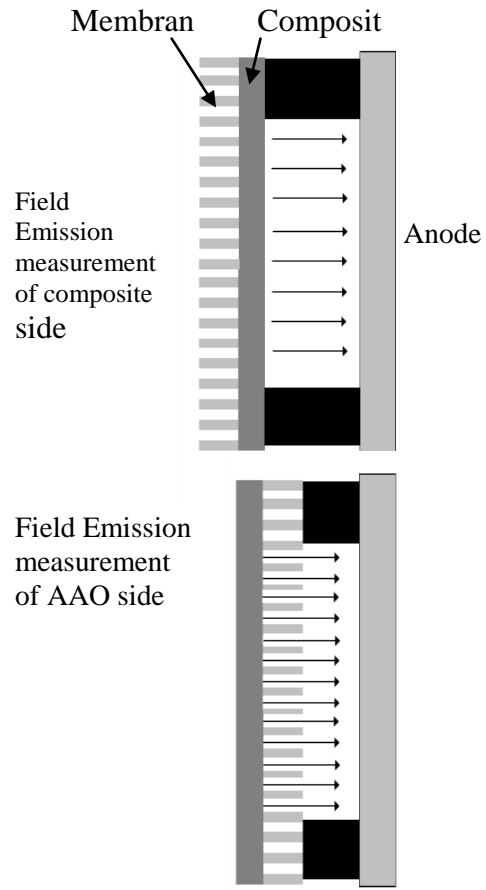

**Figure 8** The schematic view of the field emission measurement cells for testing both composite and AAO sides

publication[21].

In order to study the effect of the AAO membrane on field emission of these samples, the measurements were performed on both composite and AAO side of these samples (Figure 8). The measurement of field emission on AAO side as shown in Figure 8 has the AAO acting as a spacer, and the electrons pass through the AAO membrane en route to the other electrode. Figure 9a shows the variation of current

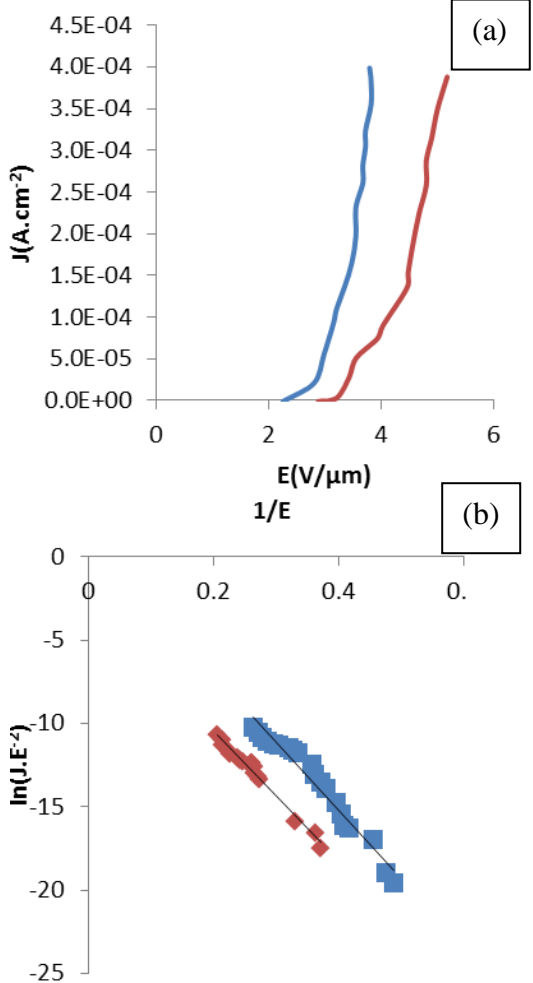

**Fig. 9** (a) Current density vs. applied field plot of composite samples on composite side (blue) and AAO membrane side (red), and (b) the Fowler-Nordheim plots developed based on the same data.

density (J) as a function applied electric field (E). By using the modified Fowler-Nordheim relationship between the applied field and the emission current[12],

$$\ln\left(\frac{J}{E^2}\right) = \ln\left(\frac{A\alpha\beta^2}{\phi}\right) + \frac{BC^2}{\sqrt{\phi}} - \frac{B\phi^{3/2}}{\beta E}$$

in which A, B and C are constants and $\phi$ is the work function (5eV for graphite), it is possible to plot $\ln(J/E^2)$ versus $1/E$ (Fig b), and use the slope to estimate mean field enhancement factor ($\beta$) as quantitative interpretation of efficiency of nanostructures to enhance the electric field and result in better field emission.

The data presented in figure 9a shows that the emission and the turn on voltage of these samples are in the same range as other electrochemically-polymerized composites.

In order to allow a comparison of these samples, and to make the effect of the AAO membrane more noticeable, the turn on field, maximum emission and field enhancement factor of these measurements are tabulated in Table 1. It can be seen that the composite side of the sample exhibits a lower turn on field, and a higher maximum emission, but the AAO side shows a slightly better mean field enhancement factor. Given that for the field emission measurement of the AAO side of the sample a much smaller surface of the sample is exposed to the electric field, it can be concluded that the AAO membrane is effectively limiting the emission by the masking parts of the emitting surface, that results in a lower current density. On the other hand, by focusing the electric field on the smaller area of sample, it serves to increase the field enhancement factor and to some extent compensate for the lower emitting area.

However, it should be noted that although this deposition technique has assisted in both sample preparation and the field emission performance, continuous operation of emission process for a long time, will result in complete degradation and removal of polypyrrole, and ultimately only a layer of nanotubes remains over the AAO membrane (Fig 10). Due to presence of the AAO membrane between anode and cathode, despite removal of the polymer binder, the membrane can prevent shorting of the system and significantly prolong the lifetime of the field emission device.

**Table 1** Field emission parameters of samples deposited on AAO membrane

| Sample | Turn on Field V/μm | Maximum Emission A/cm$^2$ | Mean field enhancement factor |
|---|---|---|---|
| Comp.-side | 2.73 | 3.53×10$^{-4}$ | 1555 |
| AAO-side | 3.34 | 2.56×10$^{-4}$ | 1762 |

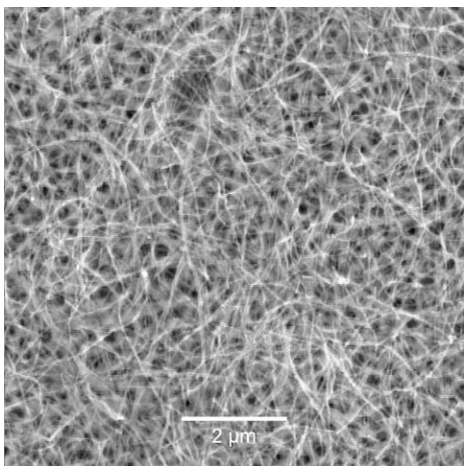

**Fig. 10** The SEM images of composite sample deposited on AAO membrane after field emission test

**Conclusion**

Herein, preparation and characterization of nanocomposites of conducting polymers (PPy and PEDOT) and carbon nanotubes or graphene and their performance as cold cathodes for field emission process is reported. These samples were prepared via chemical oxidation or electropolymerization. It was found that due to the intrinsic low stability of polypyrrole, electro- or chemically polymerized samples degrade in high emission ranges. The effect of this degradation could be observed in the field emission behavior, the vacuum of the system, and the morphology of the surface after field emission tests. Thus, although these materials can be used as field emitters, their application is limited to only low emission ranges and short operation times.

-Electropolymerization of nanocomposites can provide columnar morphology which results in better field emission, but stability is still very low.

-Despite the higher stability of PEDOT, it is not stable enough to provide a high current density.

-Electropolymerizing CNT/PPy composite directly on the surface of an AAO membrane can limit the emission surface and result in lower maximum emission and higher turn on voltage. The lower emission current and consequently lower heat generation, in addition to the possible heat dissipation by the AAO membrane can reduce the rate of the thermal degradation of polymer.

-AAO membrane can act as a physical barrier that holds nanotubes in place and allows electrons through. Thus it can prevent shorting of the field emission cell. In this way higher emission and longer life time can be achieved.